\documentclass[sigplan]{acmart}

\acmConference[PLDI’22]{ACM SIGPLAN Conference on Programming Language Design and Implementation}{June 13--17, 2022}{San Diego, CA, USA}
\acmYear{2022}
\acmISBN{} 
\acmDOI{} 
\startPage{1}

\setcopyright{rightsretained}

\usepackage{booktabs}   
\usepackage{subcaption} 
\usepackage{hyperref}
\usepackage{cleveref}
\usepackage[frozencache,cachedir=.]{minted}
\usepackage{enumitem}

\usepackage{paper-tikz-style}
\usetikzlibrary{arrows, decorations.pathmorphing}

\begin{document}

\title[On the Optimization of Equivalent Concurrent Computations]{On the Optimization of Equivalent Concurrent Computations}


\author{Henrich Lauko}
\orcid{0000-0002-5422-5884}
\affiliation{
  \institution{Trail of Bits}
  \city{New York}
  \state{NY}
  \country{USA}
}
\email{henrich.lauko@trailofbits.com}

\author{Lukáš Korenčik}
\orcid{0000-0002-1468-1594}
\affiliation{
  \institution{Trail of Bits}
  \city{New York}
  \state{NY}
  \country{USA}
}
\email{lukas.korencik@trailofbits.com}

\author{Peter Goodman}
\orcid{0000-0001-9037-5241}
\affiliation{
  \institution{Trail of Bits}
  \city{New York}
  \state{NY}
  \country{USA}
}
\email{peter@trailofbits.com}


\begin{abstract}
In this submission, we explore the use of equality saturation to optimize concurrent computations. A concurrent environment gives rise to new optimization opportunities, like extracting a common concurrent subcomputation. To our knowledge, no existing equality saturation framework allows such an optimization.
The challenge with concurrent environments is that they require non-local reasoning since parallel computations are inherently unrelated and disjoint. This submission presents a new approach to optimizing equivalent concurrent computations: extending e-graphs to capture equal concurrent computations in order to replace them with a single computation.
\end{abstract}



\maketitle

\section{Introduction}

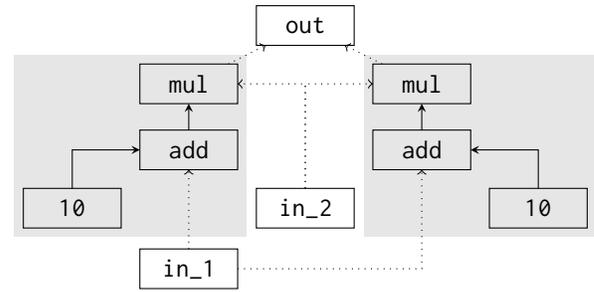
\begin{figure}[h]
    \begin{center}
    \begin{tikzpicture}[node distance=1em]
    \node [op] (out) {\texttt{out}};
    \node [op, below left = of out] (mul1) {\texttt{mul}};
    \node [op, below right = of out] (mul2) {\texttt{mul}};
    \node [op, below = of mul1] (add1) {\texttt{add}};
    \node [op, node distance = 3em, below = of add1] (in1) {\texttt{in\_1}};
    
    \node [op, below left = of add1] (con1) {\texttt{10}};
    \node [op] (in2) at (out |- con1) {\texttt{in\_2}};

    \node [op, below = of mul2] (add2) {\texttt{add}};
    \node [op, below right = of add2] (con2) {\texttt{10}};

    \begin{pgfonlayer}{background}[]
        \node[fill=black!10, fit=(mul1) (add1) (con1)] (c1) {};
    \end{pgfonlayer}
    
    \begin{pgfonlayer}{background}[]
        \node[fill=black!10, fit=(mul2) (add2) (con2)] (c2) {};
    \end{pgfonlayer}

    \draw [->, >=stealth] (add1) -- (mul1);
    \draw [->, >=stealth] (add2) -- (mul2);
    \draw [->, dotted] (mul1) -- (out);
    \draw [->, dotted] (mul2) -- (out);
    \draw [->, dotted] (in1) -- (add1);
    \draw [->, dotted] (in1) -| (add2);
    \draw [->, dotted] (in2.north) |- (mul1.east);
    \draw [->, dotted] (in2.north) |- (mul2.west);
    \draw [->, >=stealth] (con1) |- (add1);
    \draw [->, >=stealth] (con2) |- (add2);

\end{tikzpicture}
    \end{center}
    \caption{An example of an artificial arithmetic circuit containing two grayed components with equivalent concurrent executions that may be part of a larger circuit}
    \label{fig:example-circuit}
\end{figure}

We will present our approach on a real-world example of a concurrent environment -- combinational circuit \Cref{fig:example-circuit}. It is a circuit constructed from integer and bit-manipulating operations with no clock, so all units execute, regardless of whether or not they ought to.

The challenge of unoptimized combinational circuits is that they perform huge amounts of repeated work in their components. Nonetheless, our setup permits us to examine the problem of circuit optimization through a new lens: all of this repeated work can be thought of as happening in parallel. This visibility over every possible computation exposes optimization opportunities to discover the optimal sharing of sub-computations within circuit execution. An alternative way of thinking about our work is that typical sequential \textsc{cpu} circuits contain one or more hand-crafted execution units, such as arithmetic and logical units (\textsc{alu}s), and our optimization process invents shared \textsc{alu}s by observing and merging redundant computations. Our technique might also be applied to a broader set of problems, such as optimization of shared computation in threaded programs.

\begin{figure}[h]
    \centering
    \begin{tikzpicture}[node distance=0.8em]
    \node [minimum width = 3.2em, node distance=4em] (c1) {\textsf{A}};
    \node [below = of c1, op] (s1) {\texttt{comp}};
    \node [left  = of c1] (i1) {};
    \node [right = of c1] (o1) {};
    
    \begin{pgfonlayer}{background}[]
        \node[fill=black!10, fit=(c1) (s1)] (cs1) {};
    \end{pgfonlayer}

    \node [minimum width = 3.2em, node distance=4em, below = of c1] (c2) {\textsf{B}};
    \node [below = of c2, op] (s2) {\texttt{comp}};
    \node [left  = of c2] (i2) {};
    \node [right = of c2] (o2) {};
    
    \node [right = of o2] (optin) {};
    \node [node distance=3em, right = of optin] (optout) {};
    \draw [->, >=stealth, double] (optin) -- node [above,midway] {optimize} (optout);
    
    \node [right = of optout] (c2a) {$\textsf{B}_1$};
    \node [op, right = of c2a] (comp2a) {\texttt{comp}};
    \node [right = of comp2a] (c2b) {$\textsf{B}_2$};
    
    \node [left = of c2a] (ci2a) {};
    \node [right = of c2b] (co2b) {};
    
    \draw [->, >=stealth] (ci2a) -- (c2a);
    \draw [->, >=stealth] (c2b) -- (co2b);
    \draw [->, >=stealth] (c2a) -- (comp2a);
    \draw [->, >=stealth] (comp2a) -- (c2b);

    \begin{pgfonlayer}{background}[]
        \node[fill=black!10, fit=(c2) (s2)] (cs2) {};
    \end{pgfonlayer}
    
    \begin{pgfonlayer}{background}[]
        \node[fill=black!10, fit=(c2a)] (co2a) {};
    \end{pgfonlayer}
    
    \begin{pgfonlayer}{background}[]
        \node[fill=black!10, fit=(c2b)] (co2b) {};
    \end{pgfonlayer}

    \node [minimum width = 3.2em, node distance=4em, below = of c2] (c3) {\textsf{C}};
    \node [below = of c3, op] (s3) {\texttt{comp}};
    \node [left  = of c3] (i3) {};
    \node [right = of c3] (o3) {};
    
    \begin{pgfonlayer}{background}[]
        \node[fill=black!10, fit=(c3) (s3)] (cs3) {};
    \end{pgfonlayer}
    
    \node [] (c1a) at (c1 -| c2a) {$\textsf{A}_1$};
    \node [] (c1b) at (c1 -| c2b) {$\textsf{A}_2$};
    \node [left = of c1a] (ci1a) {};
    \node [right = of c1b] (co1b) {};
    \draw [->, >=stealth] (ci1a) -- (c1a);
    \draw [->, >=stealth] (c1b) -- (co1b);
    \draw [->, >=stealth] (c1a) -- (comp2a);
    \draw [->, >=stealth] (comp2a) -- (c1b);

    \begin{pgfonlayer}{background}[]
        \node[fill=black!10, fit=(c1a)] (co1a) {};
    \end{pgfonlayer}
    \begin{pgfonlayer}{background}[]
        \node[fill=black!10, fit=(c1b)] (co1b) {};
    \end{pgfonlayer}
    
    \node [] (c3a) at (c3 -| c2a) {$\textsf{C}_1$};
    \node [] (c3b) at (c3 -| c2b) {$\textsf{C}_2$};
    \begin{pgfonlayer}{background}[]
        \node[fill=black!10, fit=(c3a)] (co3a) {};
    \end{pgfonlayer}
    \begin{pgfonlayer}{background}[]
        \node[fill=black!10, fit=(c3b)] (co3b) {};
    \end{pgfonlayer}
    
    \node [left = of c3a] (ci3a) {};
    \node [right = of c3b] (co3b) {};
    \draw [->, >=stealth] (ci3a) -- (c3a);
    \draw [->, >=stealth] (c3b) -- (co3b);
    \draw [->, >=stealth] (c3a) -- (comp2a);
    \draw [->, >=stealth] (comp2a) -- (c3b);
    
    \draw [->, >=stealth] (i1) -- (c1);
    \draw [->, >=stealth] (c1) -- (o1);
    \draw [->, >=stealth] (i2) -- (c2);
    \draw [->, >=stealth] (c2) -- (o2);
    \draw [->, >=stealth] (i3) -- (c3);
    \draw [->, >=stealth] (c3) -- (o3);

\end{tikzpicture}
    \caption{The common sub-computation \texttt{comp} replaces the three concurrent components \textsf{A}, \textsf{B}, and \textsf{C}.}
    \label{fig:example-replacement}
\end{figure}
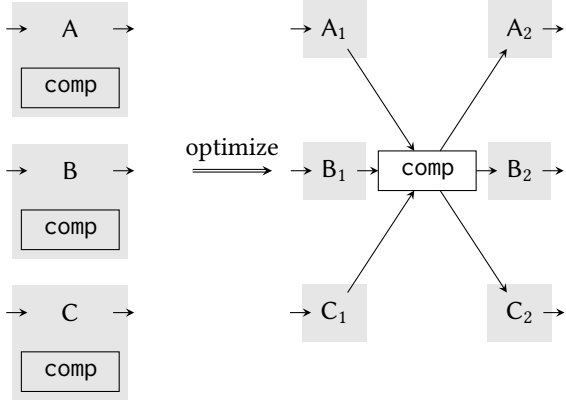

In general, our problem consists of multiple concurrent computations in an acyclic data-flow graph that share an identical subcomputation. In this case, we may want to optimize these computations by identifying where the computations overlap and replacing these areas with a single computation (see an example in \Cref{fig:example-replacement}).

The solution we propose is to extend \emph{e-graphs} \cite{nelson-egraph, nieu-egraphs} with a special bond node, so called \emph{b-node}. The \emph{b-node} serves to tie together multiple concurrent expressions. Consequently, we can unify a \emph{b-node} with our desired expression to form a single equality class and utilize a generic equality saturation algorithm. When extracting an optimal solution from the final equality graph, we can treat bond nodes as generic nodes (\emph{i.e.}, we can pick either a \emph{b-node} and all its children or any other node in the same equality class). Finally, we replace all bond nodes with adequate data-flow edges and obtain a valid circuit. We describe this process more formally in the next section.

\section{E-Graph Extension}

To be able to perform the optimization in \Cref{fig:example-replacement}, we need to allow the \emph{e-graph} to capture the information of related disjoint expressions. For this purpose, we extend the \emph{e-graphs} with bonding \emph{b-nodes}, which allow relating multiple nodes.

Informally, an extended \emph{e-graph} consists of \emph{e-classes}, as in the general case. The difference is that, in an extended \emph{e-graph}, we allow an \emph{e-class} to contain a mixed set of \emph{e-nodes} and \emph{b-nodes} at once.
As in the original definition \cite{egg}, \emph{e-nodes} represent terms of a modeled language, and \emph{b-nodes} represent semantically bonded nodes. The \emph{b-node} allows one to replace all bonded nodes at once because one can treat all bonded nodes as a single node. More formally, \emph{e-nodes} and \emph{b-nodes} are defined as follows:
\begin{itemize}
    \item An \emph{e-node} is a function symbol paired with a list of children \emph{e-classes}.
    \item A \emph{b-node} is a unique symbol that also keeps a \emph{bond-map}, a mapping between its parent and children \emph{e-classes}.
\end{itemize}

\noindent
Two \emph{b-nodes} are considered equal when their \emph{bond-maps} are identical.


This representation allows us to capture in an \emph{e-graph} that some \emph{e-classes} are related (\emph{e.g.}, in our case, they perform the same computation concurrently).

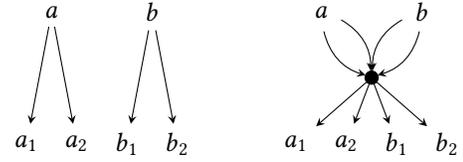
\begin{figure}[h]
    \centering
    \begin{tikzpicture}[node distance=0.4em]
    \node [common] (al) {$a_1$};
    \node [common, right = of al] (ar) {$a_2$};
    
    \coordinate (ma) at ($(al)!0.5!(ar)$);
    \node (a) [above of = ma, node distance = 5em] {$a$};
    
    \node [common, right = of ar] (bl) {$b_1$};
    \node [common, right = of bl] (br) {$b_2$};
    
    \coordinate (mb) at ($(bl)!0.5!(br)$);
    \node (b) [above of = mb, node distance = 5em] {$b$};
    
    \node [common, node distance=3em, right = of br] (alb) {$a_1$};
    \node [common, right = of alb] (arb) {$a_2$};
    \node [common, right = of arb] (blb) {$b_1$};
    \node [common, right = of blb] (brb) {$b_2$};
    
    \coordinate (m) at ($(alb)!0.5!(brb)$);
    \node (bond) [above of = m, node distance = 2.5em, circle, draw, fill=black, minimum size=0.5em, inner sep=0pt, outer sep=0pt] {};
    
    \coordinate (mab) at ($(alb)!0.5!(arb)$);
    \node (ab) [common, above of = mab, node distance = 5em] {$a$};
    
    \coordinate (mbb) at ($(blb)!0.5!(brb)$);
    \node (bb) [common, above of = mbb, node distance = 5em] {$b$};
    
    \draw [->, >=stealth] (a) -- (al);
    \draw [->, >=stealth] (a) -- (ar);
    
    \draw [->, >=stealth] (b) -- (bl);
    \draw [->, >=stealth] (b) -- (br);
    
    \draw [->, >=stealth] (ab)  to[bend left] (bond);
    \draw [->, >=stealth] (ab) to[bend right] (bond);
    
    \draw [->, >=stealth] (bb)  to[bend left] (bond);
    \draw [->, >=stealth] (bb) to[bend right] (bond);
    
    \draw [->, >=stealth] (bond) -- (alb);
    \draw [->, >=stealth] (bond) -- (arb);
    
    \draw [->, >=stealth] (bond) -- (blb);
    \draw [->, >=stealth] (bond) -- (brb);

\end{tikzpicture}
    \caption{Nodes $a$ and $b$ on the left are bonded by the \emph{b-node} on the right, denoted by the center black node}
    \label{fig:bond-example}
\end{figure}

Imagine that we have three \emph{e-classes} denoted as $p_i$, each with two children $c_{ij}$. To relate these \emph{e-classes} via a bond, we store the \emph{bond-map}
$
    [p_1 \rightarrow \{c_{11}, c_{12}\},\, p_3 \rightarrow \{c_{21}, c_{22}\},\, p_2 \rightarrow \{c_{31}, c_{32}\},]
$
The corresponding \emph{e-graph} with bonded nodes is depicted in \Cref{fig:bond-example}.

We store the \emph{bond-map} so that the \emph{b-node} can be removed after the equality saturation process finishes. It might happen that a \emph{b-node} is an optimal representation of an \emph{e-class}. However, we do not want to preserve \emph{b-nodes} in the final solution. Therefore, we utilize their \emph{bond-maps} to disperse the \emph{b-nodes} by linking parents with their children from the \emph{bond-maps}. In other words, dispersion is an inverse operation to bonding.

\begin{example}
In our circuit use case, we aim to extract shared arithmetic computations (\emph{e.g.}, multiplications or additions). This use case is unique in how it treats operation inputs; we can synthesize a special \emph{advice} input that essentially behaves like any input from concurrent environments.
Therefore, we can bond all concurrent multiplications regardless of their inputs and unify the \emph{b-node} with a single multiplication with \emph{advice} in place of all inputs. This optimization can be viewed as extracting an \textsc{alu} for a particular operation in the circuit. Our rewrite patterns for multiplication optimization are as follows:
\begin{enumerate}[leftmargin=*,align=left]
\item First, we perform upcasting of all multiplications with various bitwidths, \texttt{bw}, to the largest bitwidth of 64 bits:
\begin{minted}[fontsize=\small]{lisp}
(mul:bw ?a ?b) => 
(trunc:bw (mul:64 (zext:64 ?a) (zext:64 ?b)))
\end{minted}
\smallskip

\item Then, we gather all 64-bit multiplications and bond them:
\begin{minted}[fontsize=\small]{lisp}
(let Muls (mul:64)...) => (let Bond (bond Muls...))
\end{minted}
\smallskip

\item Lastly, we unify the bonded nodes with a single replacement multiplication 
that takes \texttt{advice} values as input:
\begin{minted}[fontsize=\small]{lisp}
(unify Bond (mul:64 advice:64 advice:64))
\end{minted}

\end{enumerate}
\end{example}

\section{Limitations}

Unfortunately, bonding is sensitive to dependencies of rewrite rules. Suppose you have a rule that generates a new possibility to extend a bond set every time. If such a rule is interleaved with a bonding rule, it may cause an infinite chain of \emph{b-nodes}. To mitigate this issue, we utilize a suboptimal solution that bonds nodes only before or after generic equality saturation.

Another challenging problem arises when candidates for bonding need to satisfy some constraints. In such a case, one needs to find a maximal satisfiable set or generate a combinatorial number of \emph{b-nodes}.
Again, we have opted for a suboptimal solution that eagerly finds some possible set to mitigate this problem.

\section{Conclusion}
In this submission, we presented ongoing work on concurrency optimization. We challenge the locality of equality optimization and introduce a new kind of \emph{e-graph} node, called the \emph{b-node}, to reason about disjoint computations. This allows us to apply equality saturation to a new set of problems, especially the extraction of common subcomputations in a concurrent environment. We discussed new challenges introduced by node bonding and its relationship to traditional equality saturation. We demonstrated our approach using a real example of circuit optimization.

\section{Acknowledgement}
This research was developed with funding
from the Defense Advanced Research Projects Agency (DARPA).

The views, opinions and/or findings
expressed are those of the author and should not be interpreted as
representing the official views or policies of the Department of Defense or
the U.S. Government.

\section*{DISTRIBUTION STATEMENT A} Approved for public release, distribution unlimited.
\bibliography{main}



\end{document}